\documentclass[fleqn,10pt,twocolumn]{main}

\usepackage{amsmath, amssymb, amsfonts}
\usepackage{enumitem}
\setlist[itemize]{left=1.5em}  
\usepackage{textcomp} 
\usepackage{algorithm}
\usepackage{subfig}
\usepackage{bm}
\usepackage{algpseudocode}

\newcounter{remark}
\newenvironment{remark}{%
  \refstepcounter{remark}\par\noindent\textbf{Remark~\theremark.}\ }{\par}

\makeatletter
\renewcommand\section{\@startsection{section}{1}{\z@}
  {-2.5ex \@plus -1ex \@minus -.2ex}
  {1.5ex \@plus .2ex}
  {\centering\normalfont\Large\bfseries}}

\renewcommand\subsection{\@startsection{subsection}{2}{\z@}
  {-2.25ex \@plus -1ex \@minus -.2ex}
  {1ex \@plus .2ex}
  {\normalfont\normalsize\bfseries}}

\renewcommand\subsubsection{\@startsection{subsubsection}{3}{\z@}
  {-2ex \@plus -1ex \@minus -.2ex}
  {1ex \@plus .2ex}
  {\normalfont\normalsize\itshape}}
\makeatother

\begin{document}

\twocolumn[
  \vspace*{2.0em} 
  \begin{center}
    {\Large\bfseries Autonomous Detection and Coverage of Unknown Target Areas by Multi-Agent Systems\par}
    \vspace{0.5em}
    {\large Jie Song${}^{1\dagger}$, Yang Bai${}^{2}$, Mikhail Svinin${}^{3}$, Naoki Wakamiya${}^{1}$\par}
    \vspace{0.5em}
    {\normalsize
      ${}^{1}$Department of Bioinformatic Engineering, University of Osaka, Osaka, Japan\\
      (E-mail: j-song@ist.osaka-u.ac.jp; wakamiya@ist.osaka-u.ac.jp)\\
      ${}^{2}$Graduate School of Advanced Science and Engineering, Hiroshima University, Hiroshima, Japan\\
      (E-mail: yangbai@hiroshima-u.ac.jp)\\
      ${}^{3}$Graduate School of Information Science and Engineering, Ritsumeikan University, Osaka, Japan\\
      (E-mail: svinin@fc.ritsumei.ac.jp)
    \par}
  \end{center}

  \noindent\vspace{1.0em}{\normalsize\textbf{Abstract:} \normalfont
  This paper presents a novel coverage control algorithm for multi-agent systems, where each agent has no prior knowledge of the specific region to be covered. The proposed method enables agents to autonomously detect the target area and collaboratively achieve full coverage. Once an agent detects a part of the target region within its sensor range, a dynamically constructed density function is generated to attract nearby agents. By integrating this density-driven mechanism with Centroidal Voronoi Tessellation (CVT), the agents are guided to achieve optimal spatial distribution. Additionally, Control Barrier Functions (CBFs) are employed to ensure collision avoidance and maintain non-overlapping sensor coverage, enhancing both safety and efficiency. Simulation results verify that agents can independently locate and effectively cover the target area.\par}

  \noindent\vspace{1.0em}{\normalsize\textbf{Keywords:} \normalfont
  multi-agent systems; coverage control\par}

  \vspace{1.4em} 
]

\pagestyle{headings}    


\section{Introduction}
In recent years, the problem of multi-agent coverage has attracted increasing attention, particularly in applications such as cooperative transportation and environment monitoring \cite{Bai2022,Shibata2021,Chen2024,Bulla2021,Sacco2021,Dobrokvashina2021,Bai_23,Guo2021,Liu2025,Bai2018,Bai2025,magid2022asia}. Compared to single-agent systems, multi-agent systems offer higher efficiency and greater robustness, making them more suitable for complex and dynamic environments. However, many existing multi-agent coverage algorithms assume that the target area is known beforehand. Each agent is given prior information about the region, including its shape and location. This strong dependency on prior knowledge limits the flexibility and applicability of these algorithms in real-world scenarios. Therefore, this paper aims to develop a new multi-agent coverage algorithm for unknown environments. In such environments, agents need to explore and adapt to the target area without prior information.

Several studies have explored the multi-agent coverage problem in unknown environments, aiming to improve system autonomy and adaptability. For instance, Shi et al. \cite{Shi2018} proposed a cooperative sweep coverage algorithm capable of handling unknown and irregular-shaped areas. Their method utilizes virtual elastic forces to maintain formation and ensure full coverage. Similarly, Rastgoftar \cite{Rastgoftar2025} introduced a decentralized deep learning-based control architecture that enables a multi-agent team to cover an unknown target without relying on prior map information. These approaches demonstrate the potential of deploying multiple agents in unknown and unstructured environments.

However, many existing algorithms are primarily designed for single and continuous target regions. For example, the method proposed in \cite{Xiao2020} transforms the continuous coverage process into a discrete traversal of a single target area, while the dynamic reinforcement learning scheme in \cite{Cao2024} focuses on tracking and covering one dynamic region in a completely unknown environment. Consequently, these algorithms lack the ability to concurrently address multiple disconnected or spatially scattered target areas. This limitation significantly reduces their applicability in complex real-world scenarios where multiple, spatially separated regions need to be monitored or searched simultaneously.

To address the challenge of covering multiple disconnected target areas in unknown environments, researchers have increasingly turned to heuristic-based methods \cite{Pan2022,Shao2024,Shao2025,Zhang2024,Zhang2025}. These methods offer flexible and computationally efficient solutions by employing domain knowledge, empirical strategies, or approximation techniques to guide decision-making in complex and uncertain settings. Unlike model-based approaches that require full environmental knowledge, heuristic-based methods rely on locally available information to generate effective, if not optimal, coverage behaviors.

One of the most prominent heuristic strategies is the artificial potential fields (APF), which models the environment as a synthetic field comprising attractive forces from target areas and repulsive forces from obstacles \cite{Pan2019,Matoui2019,Tallamraju2018,Hamed2022,Lu2025}. Agents move in response to the gradient of the potential field, enabling real-time adaptation to environmental changes. Zheng et al. \cite{Zheng2015} introduced a hybrid approach combining hierarchical reinforcement learning with artificial potential fields to improve convergence speed and coverage performance. Similarly, S. Ivić et al. \cite{Ivic2016} developed an ergodic control strategy embedded in a potential field framework to enable agents to cooperatively explore and cover dynamic and unknown regions.

Other studies have extended this concept further. Qin et al. \cite{Qin2023} employed a region-division strategy combined with Dijkstra-based heuristics to improve path smoothness and subregion coherence in multi-agent coverage path planning. Meanwhile, in the context of pursuit-evasion games, David Wiman and David Lindgren \cite{Wiman2025} used frontier-based exploration. This heuristic method encourages agents to move toward unknown boundaries to track adversarial intruders in unknown terrain. Similarly, F. Gul et al. \cite{Gul2022} explored exploration and coverage through a centralized strategy using geodesic partitioning to efficiently allocate tasks across agents. Despite their effectiveness in large-scale and real-time scenarios, heuristic-based methods such as PFM often suffer from a key limitation: the lack of global optimality guarantees. Agents may become trapped in local minima due to conflicting virtual forces, particularly in cluttered or narrow environments. 

To enable efficient multi-agent coverage of multiple unknown target areas while avoiding local minima, we propose a novel approach that integrates CVT with a density function. By continuously updating the density distribution through a centralized controller based on sensor-detected information and iteratively adjusting agent positions via CVT, our method ensures that each agent converges toward an optimal coverage configuration. This mechanism allows the team to identify and collectively cover all target regions, even when their number, shape, and location are initially unknown. In addition, we incorporate CBFs into the control framework to ensure collision avoidance among agents during the coverage process.

This paper is organized as follows: In Section~\ref{Preliminary}, some preliminaries about CVT are introduced. Section~\ref{Controller design} presents the controller design, including the integration of CVT and the construction of the density function. Simulation results demonstrating the effectiveness of the proposed method are provided in Section~\ref{simulation}, and conclusions are drawn in Section~\ref{conclusion}.

\section{Preliminary} \label{Preliminary}
To address the problem of covering unknown target areas, we propose a multi-agent control strategy that enables cooperative coverage using multiple agents. Each agent is modeled as a mass point and is capable of perceiving a limited portion of the target areas. A CVT-based method is employed as the core technique to guide the coverage process.


In this study, we define $\mathcal{D} \subset \mathbb{R}^2$ as a two-dimensional domain representing the target areas to be covered by $N$ agents. Let \(\mathcal{N} = \{1, \dots, N\}\) be the index set of all agents. Let $\boldsymbol{p} = \{ \boldsymbol{p}_i \in \mathbb{R}^2 \mid i \in \mathcal{N} \}$ denote the collection of position vectors of all $N$ agents. A point $\boldsymbol{q} \in \mathcal{D}$ represents an arbitrary location within the target areas. The Voronoi tessellation over the domain is given by:
\begin{equation}\label{voronoi}
V_{i}\left ( \boldsymbol{p} \right )= \left \{ \boldsymbol{q} \in \mathcal{D} \mid \left \| \boldsymbol{q} - \boldsymbol{p}_i \right \| \leq \left \| \boldsymbol{q} - \boldsymbol{p}_j \right \|,\, \forall j \neq i \right \},
\end{equation}

ensuring that each agent $i$ is associated with a specific Voronoi subregion $V_i$.

To evaluate the effectiveness of the agent distribution, we define the locational cost function as:
\begin{equation}\label{H0}
\mathcal{H}(\boldsymbol{p}, t) = \sum_{i=1}^{N} \int_{V_i} \left \| \boldsymbol{q} - \boldsymbol{p}_i \right \|^2 \, \phi(\boldsymbol{q}, t) \, \mathrm{d} \boldsymbol{q},
\end{equation}
where $\phi(\boldsymbol{q}, t)$ is a density function, assumed to be strictly positive and bounded over the domain $\mathcal{D}$. This function represents the relative importance of different spatial locations and guides agents to concentrate in higher-density regions.

To achieve an optimal coverage distribution within the domain $\mathcal{D}$ at any time $t$, the agent positions $\boldsymbol{p}$ should be adjusted to minimize the locational cost function $\mathcal{H}$. According to \cite{Cortes_04}, a necessary condition for optimality is:
\begin{equation} \label{E6}
\boldsymbol{p}_i(t) = \boldsymbol{c}_i(\boldsymbol{p}, t), \quad i \in \mathcal{N},
\end{equation}
where $\boldsymbol{c}_i$ is the centroid of the Voronoi cell $V_i$ computed with respect to the density function as defined in \cite[Eq.~(3)]{Bai2022}.

\begin{remark}
\label{optimal}
When condition~\eqref{E6} is satisfied, the multi-agent system reaches a locally optimal configuration, where each agent is located at the centroid of its corresponding Voronoi cell. This configuration defines a CVT, which minimizes the coverage cost function under the given density distribution.
\end{remark}

\section{Coverage strategy and controller design} \label{Controller design}
In this section, we present the overall framework for multi-agent detection and coverage. First, a density-based strategy is introduced to guide agents toward the detected areas of interest. Then, a composite controller integrating CVT and CBF is developed to ensure both efficient coverage and collision avoidance.

\subsection{Target area detection and coverage}
The problem of detecting and covering unknown regions can be formulated as a coverage control problem, in which multiple agents are deployed over a workspace to achieve effective spatial coverage. Unlike conventional coverage settings, a distinguishing feature of this scenario is that the spatial distribution of high-priority regions is initially unknown to the agents.

To address this challenge, we propose a new a time-varying density function, denoted by $\phi(\boldsymbol{q}, t)$. By employing the newly designed density function, agents can adaptively adjust their estimation of important regions and ultimately achieve optimal coverage of the entire area.

The form of $\phi(\boldsymbol{q}, t)$ is as follows
\begin{align}
\label{eq6}
\phi(\boldsymbol{q}, t) =\ & \phi_0 + \sum_{i=1}^{N} \rho_i(t) \frac{\omega_i}{2\pi \sigma_{x_i} \sigma_{y_i}} \notag \\
& \exp\left( -\frac{1}{2} \left( 
\frac{(q_x - \mu_{x_i})^2}{\sigma_{x_i}^2} + 
\frac{(q_y - \mu_{y_i})^2}{\sigma_{y_i}^2} 
\right) \right),
\end{align}
where $\phi_{0} > 0$ is a small positive constant that ensures the density function remains strictly positive throughout the domain. For each agent $i \in \mathcal{N}$, the parameters $\omega_i > 0$, $\sigma_{x_i} > 0$, and $\sigma_{y_i} > 0$ represent the weight and the standard deviations of the $i$-th Gaussian component along the $q_x$ and $q_y$ axes, respectively. The mean values $\mu_{x_i}$ and $\mu_{y_i}$ specify the center of the $i$-th Gaussian component, typically corresponding to the current position of agent $i$. The vector $\boldsymbol{q} = [q_x, q_y]^{\top}$ denotes a generic point in the domain $\mathcal{D}$. In practice, once an important region is identified through sensor detection, $\boldsymbol{q}$ represents any candidate point within this region when evaluating the density function and computing Voronoi centroids. The coordinates $(q_x, q_y)$ are sampled from the detected region during numerical integration, and the density peaks are centered at $(\mu_{x_i}, \mu_{y_i})$ corresponding to the positions of agents that have detected the target area. This formulation allows the CVT computation to distribute agents according to the actual shape and location of the detected region, rather than forcing them to converge to its geometric center, thereby supporting target regions of arbitrary shapes.

The binary switching variable $\rho_i(t) \in \{0, 1\}$ indicates whether agent $i$ currently detects a significant region. It is defined as
\begin{equation}
\label{eq:rho}
\rho_i(t) =
\begin{cases}
1, & \text{if agent $i$ detects a target area at time $t$}, \\
0, & \text{otherwise}.
\end{cases}
\end{equation}
This detection-dependent activation mechanism dynamically modulates the density function, enabling the coverage control scheme to adapt to the evolving distribution of high-priority regions.

With the newly designed event-triggered density function, the overall procedure unfolds as follows, from the initial deployment of agents to the final coverage configuration.

At the beginning, agents are uniformly distributed across the workspace, without any prior knowledge of the location or shape of the target regions. If no agent detects a target within its sensor range, the agents gradually increase their mutual distances. This results in a more dispersed formation, enabling the team to explore a wider area more efficiently.

When an agent detects a target area within its sensor range, the detection status $\rho_i(t)$ in Equation ~\eqref{eq:rho} is set to $1$, which updates $\phi(\boldsymbol{q}, t)$ defined in Equation (4). Each peak of the density corresponds to the current position of an agent that detects a target area.

Using the updated density, a Voronoi tessellation is calculated as in Equation (1). Each agent then computes the centroid of its Voronoi region with respect to the current density, following Equation (4). The goal is to minimize the locational cost function defined in Equation (2), which leads agents to move toward regions with higher density.

As agents move toward their respective centroids, more agents may detect target regions. New peaks are added to the density function at their positions, resulting in a multimodal distribution that reflects the evolving detection landscape. Through this iterative process, the system gradually converges to a configuration that ensures comprehensive coverage of all significant regions.

\subsection{Controller design}

This subsection presents the detailed formulation of the proposed composite motion controller, which combines CVT-based positioning with CBF-based collision avoidance.

To achieve effective and safe coverage of the target areas, we design a composite motion controller that combines CVT-based optimal positioning with a CBF to ensure inter-agent collision avoidance.

Each agent is modeled as a first-order integrator:
\begin{equation}\label{dynamics}
    \dot{\boldsymbol{p}}_{i}(t) = \boldsymbol{u}_{i}(t),
\end{equation}
where $\boldsymbol{u}_i \in \mathbb{R}^2$ denotes the control input of agent $i$.

Following the CVT principle, the nominal control input is defined as:
\begin{equation}
    \boldsymbol{u}_{i}^{\mathrm{nom}}(t) = -k(\boldsymbol{p}_{i}(t) - \boldsymbol{c}_{i}(t)),
\end{equation}
where $\boldsymbol{c}_i(t)$ is the centroid of the Voronoi region $V_i$ computed with respect to a density function $\phi(\boldsymbol{q}, t)$~\cite{Cortes_04}, and $k > 0$ is a positive gain.

To guarantee safety, we incorporate CBFs into the control design. For each pair of agents $(i,j)\in \mathcal{N} \times \mathcal{N}$, $i\ne j$, we define the following safety set:
\begin{equation}
\begin{aligned}
\mathcal{C}_{ij}(t) = \bigg\{\, i \in \mathcal{N} \ \bigg|\ 
& \|\boldsymbol{p}_i(t) - \boldsymbol{p}_j(t)\|^2 
  - d_{\mathrm{min}}^2 \geq 0,  
\\
& \qquad\qquad\text{for some } j \ne i \,\bigg\}
\end{aligned}
\end{equation}
where $d_{\mathrm{min}}\in \mathbb{R}$ is the minimum safety distance.

A zeroing CBF $h_{ij}(t) = \|\boldsymbol{p}_i(t) - \boldsymbol{p}_j(t)\|^2 - d_{\mathrm{min}}^2$  (see~\cite{Ames2017}) ensures that the constraint $h_{ij}(t) \geq 0$ is forward invariant if the following condition is satisfied:
\begin{equation}
    \dot{h}_{ij}(t) + \gamma h_{ij}(t) \geq 0,
\end{equation}
 for all $t\geq 0$, where $\gamma > 0$ is a positive constant. This condition is enforced by solving a quadratic program (QP) at each time step:

\begin{align}
    \min_{\boldsymbol{u}_i} \quad & \|\boldsymbol{u}_i(t) - \boldsymbol{u}_i^{\mathrm{nom}}(t)\|^2 \notag \\
    \text{s.t.} \quad & \frac{\partial h_{ij}(t)}{\partial \boldsymbol{p}_i(t)} \boldsymbol{u}_i(t) 
    + \frac{\partial h_{ij}(t)}{\partial \boldsymbol{p}_j(t)} \boldsymbol{u}_j(t) \notag  + \gamma h_{ij}(t)\\
    & \geq 0, \quad
    \text{for all } (i,j) \in \mathcal{N} \times \mathcal{N},\ i \ne j.
\end{align}
Through this formulation, each agent moves toward its Voronoi centroid to minimize the locational cost while guaranteeing collision-free motion with respect to its neighbors. As agents dynamically update their positions and sensing information, the Voronoi partitions and density functions evolve, enabling safe coverage of the AOI.

Overall, this composite controller integrates coverage optimization via CVT with formal safety guarantees through CBFs, forming the complete control law used in our multi-agent coverage strategy.

\section{Simulation results}\label{simulation}
\begin{figure}[htbp]
    \centering
    \subfloat[]{\includegraphics[width=0.50\textwidth]{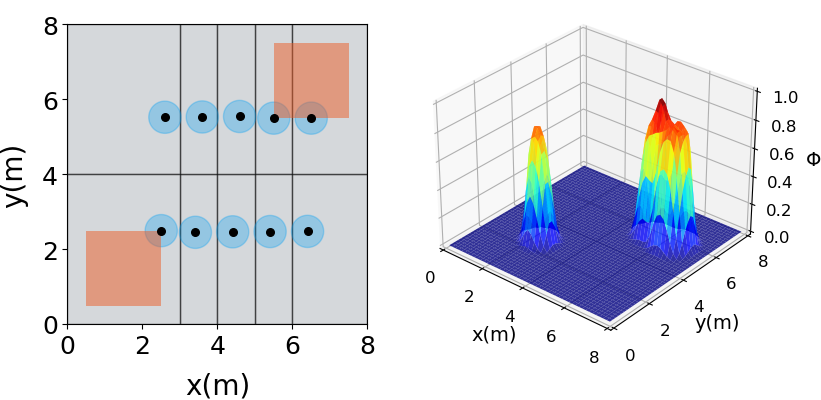}}
    \hfill
    \subfloat[]{\includegraphics[width=0.50\textwidth]{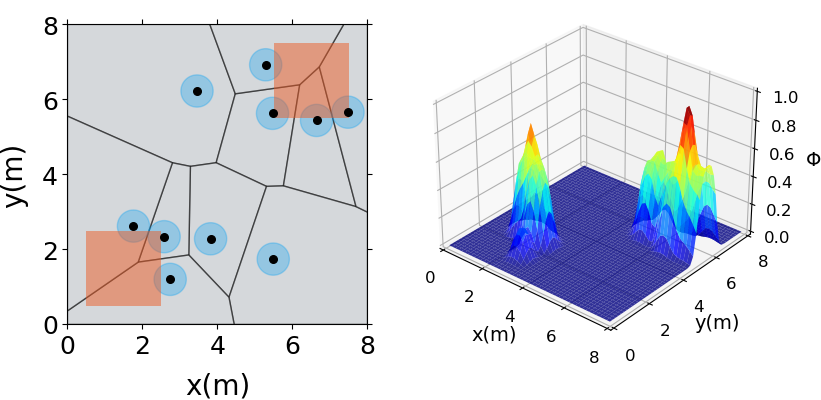}}\\
    \subfloat[]{\includegraphics[width=0.50\textwidth]{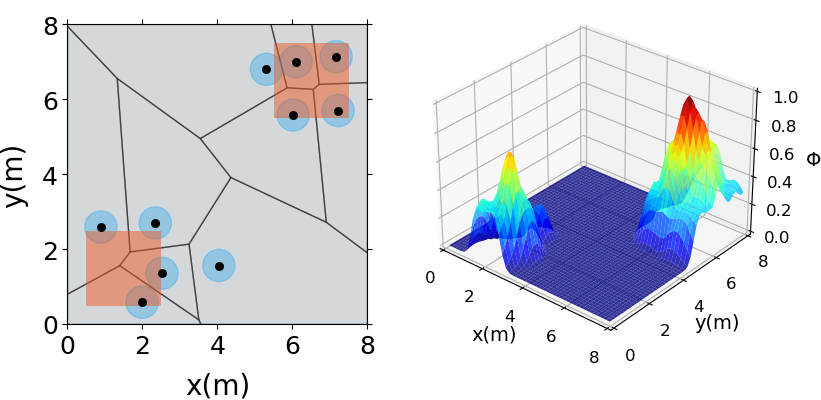}}
    \hfill
    \subfloat[]{\includegraphics[width=0.50\textwidth]{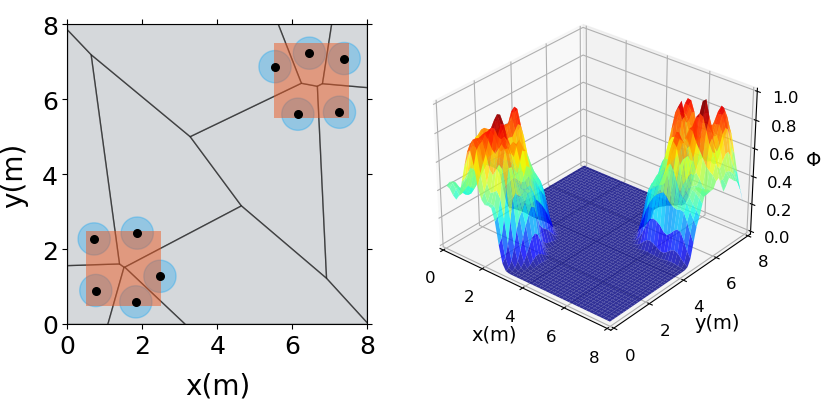}}
    \caption{Snapshots of the simulation process at different time steps: (a) $t = 0\,\mathrm{s}$, (b) $t = 15\,\mathrm{s}$, (c) $t = 26\,\mathrm{s}$, and (d) $t = 64\,\mathrm{s}$.
}
    \label{fig:coverage_snapshots_2AOI}
\end{figure}

\begin{figure}[htbp]
    \centering
    \includegraphics[width=0.45\textwidth]{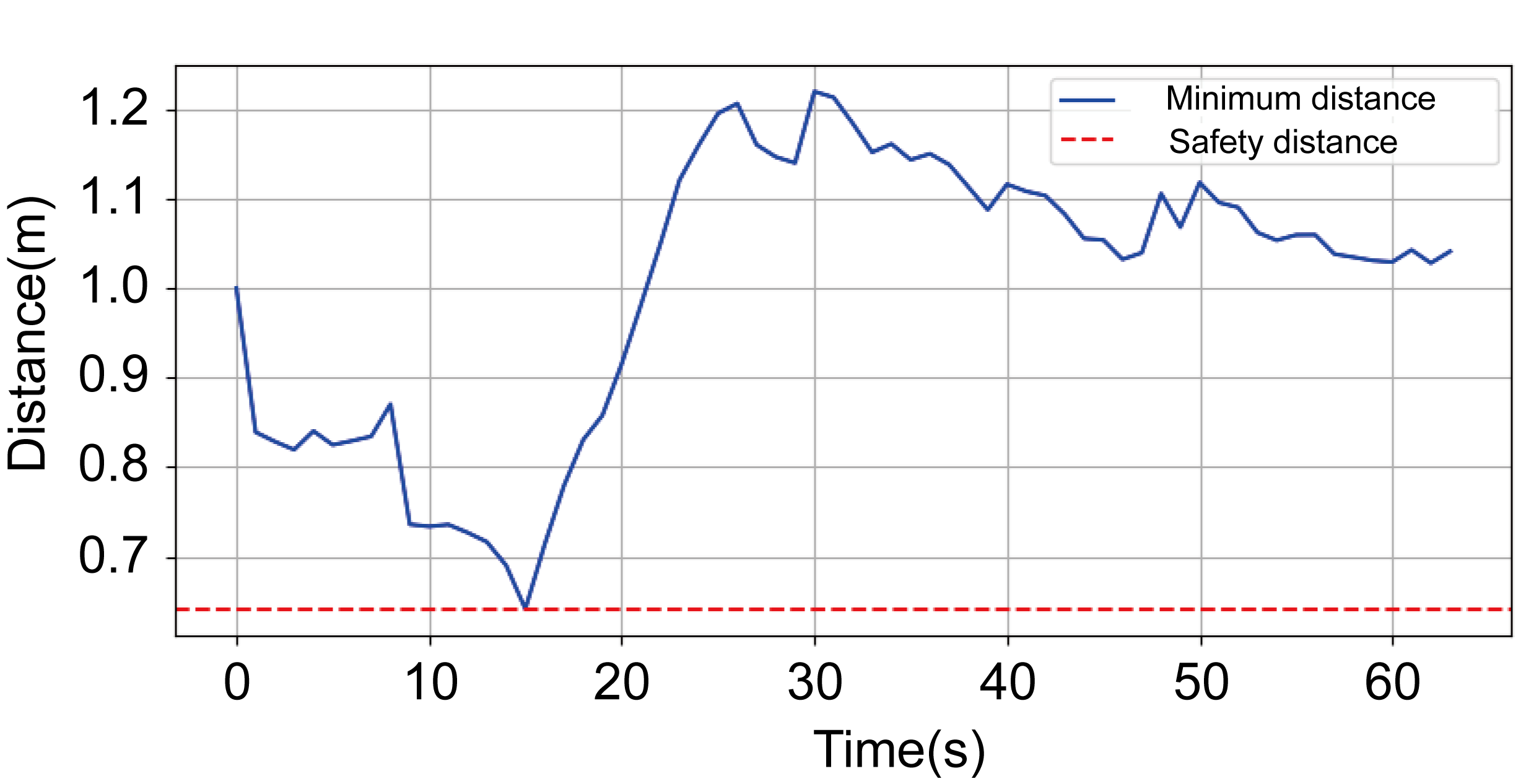}
    \caption{Minimum pairwise distance between agents over time.}
    \label{fig:minimum_distance_2AOI}
\end{figure}

\begin{figure}[htbp]
    \centering
    \includegraphics[width=0.45\textwidth]{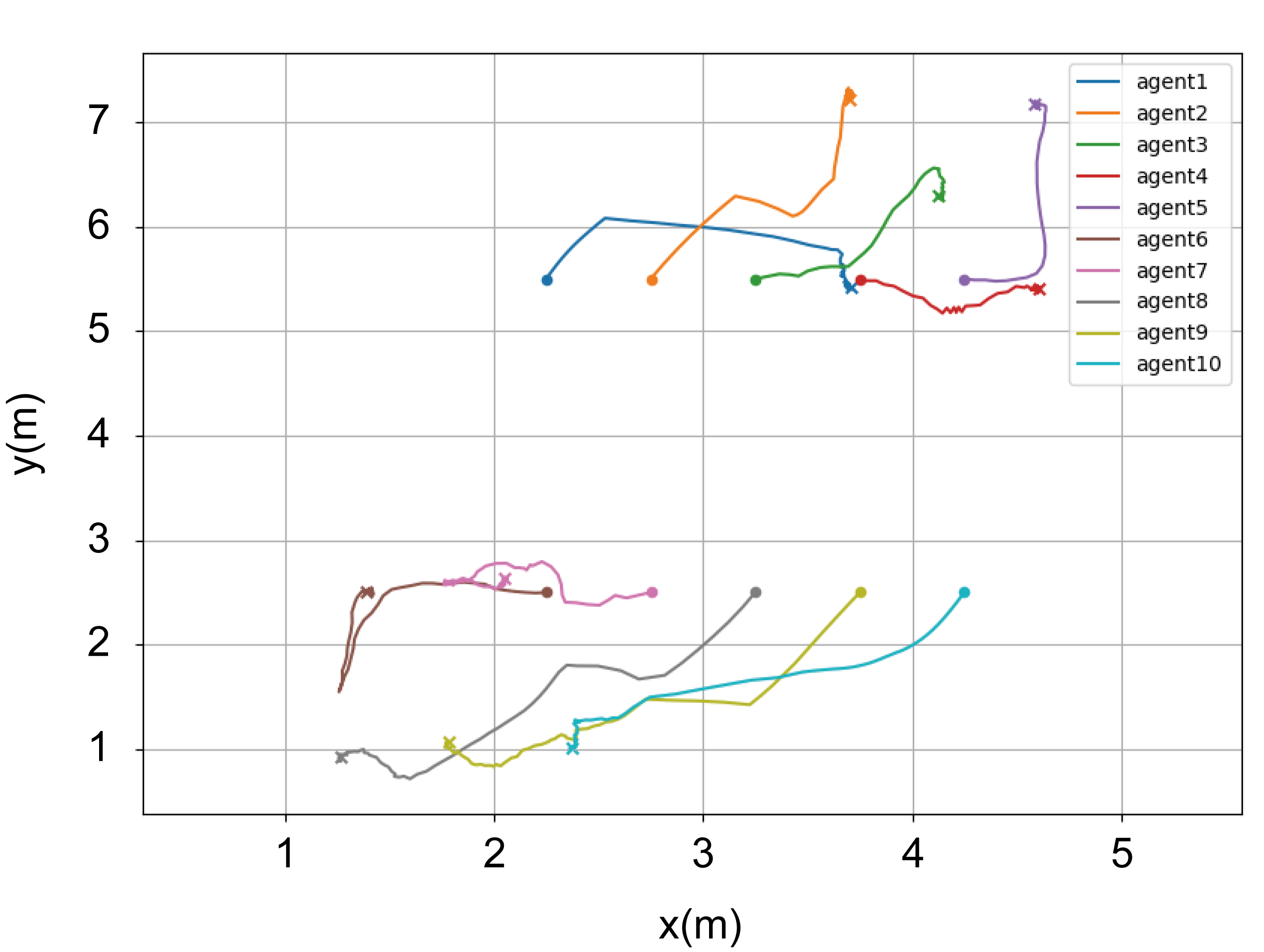}
    \caption{Complete agent trajectories during the simulation.}
    \label{fig:agent_paths_2AOI}
\end{figure}

\begin{figure}[htbp]
    \centering
    \subfloat[]{\includegraphics[width=0.50\textwidth]{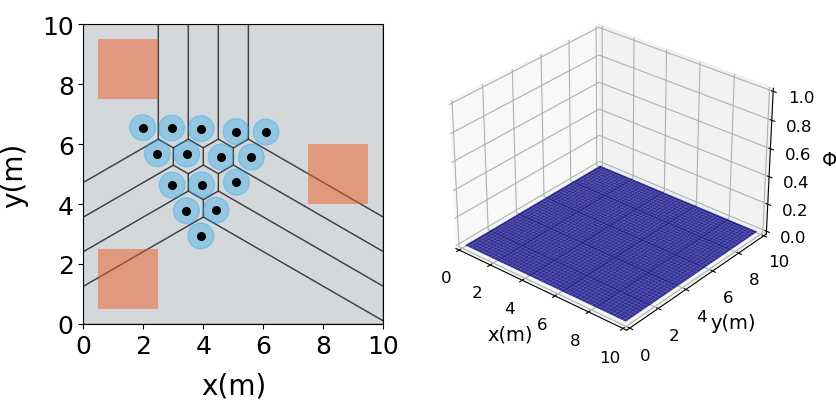}}
    \hfill
    \subfloat[]{\includegraphics[width=0.50\textwidth]{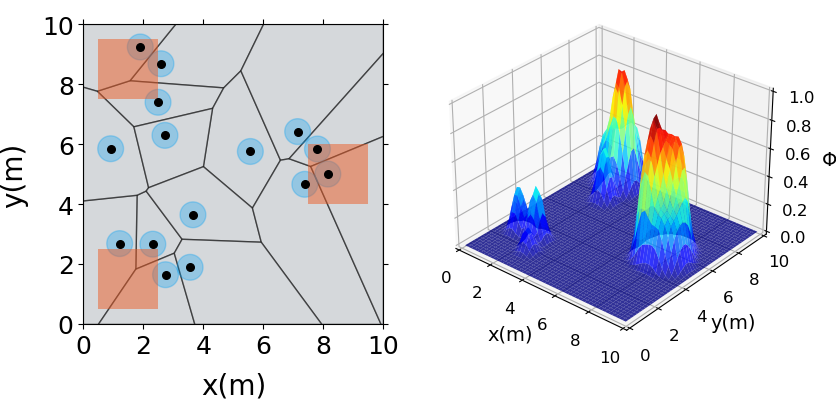}}\\
    \subfloat[]{\includegraphics[width=0.50\textwidth]{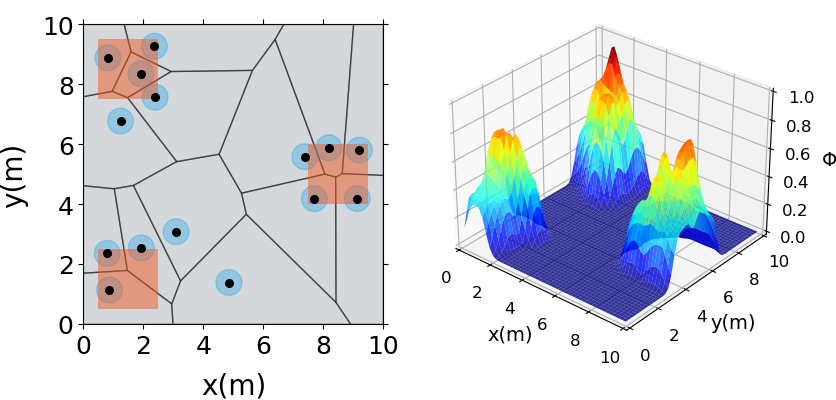}}
    \hfill
    \subfloat[]{\includegraphics[width=0.50\textwidth]{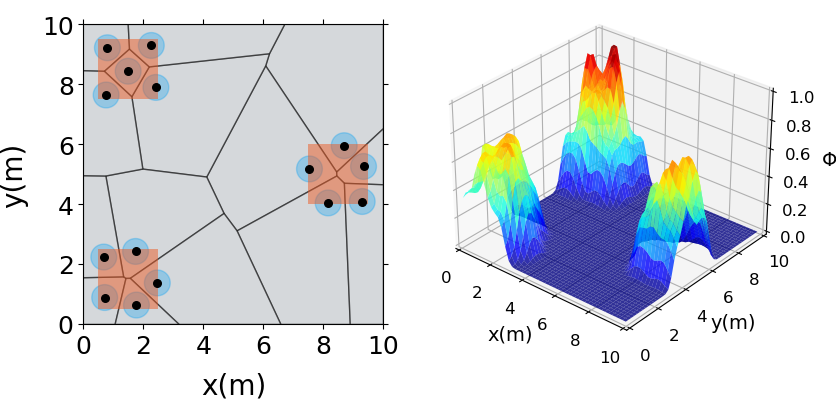}}
    \caption{Snapshots of the simulation process at different time steps: (a) $t = 0\,\mathrm{s}$, (b) $t = 12\,\mathrm{s}$, (c) $t = 32\,\mathrm{s}$, and (d) $t = 64\,\mathrm{s}$.
}
    \label{fig:coverage_snapshots_3AOI}
\end{figure}

\begin{figure}[htbp]
    \centering
    \includegraphics[width=0.45\textwidth]{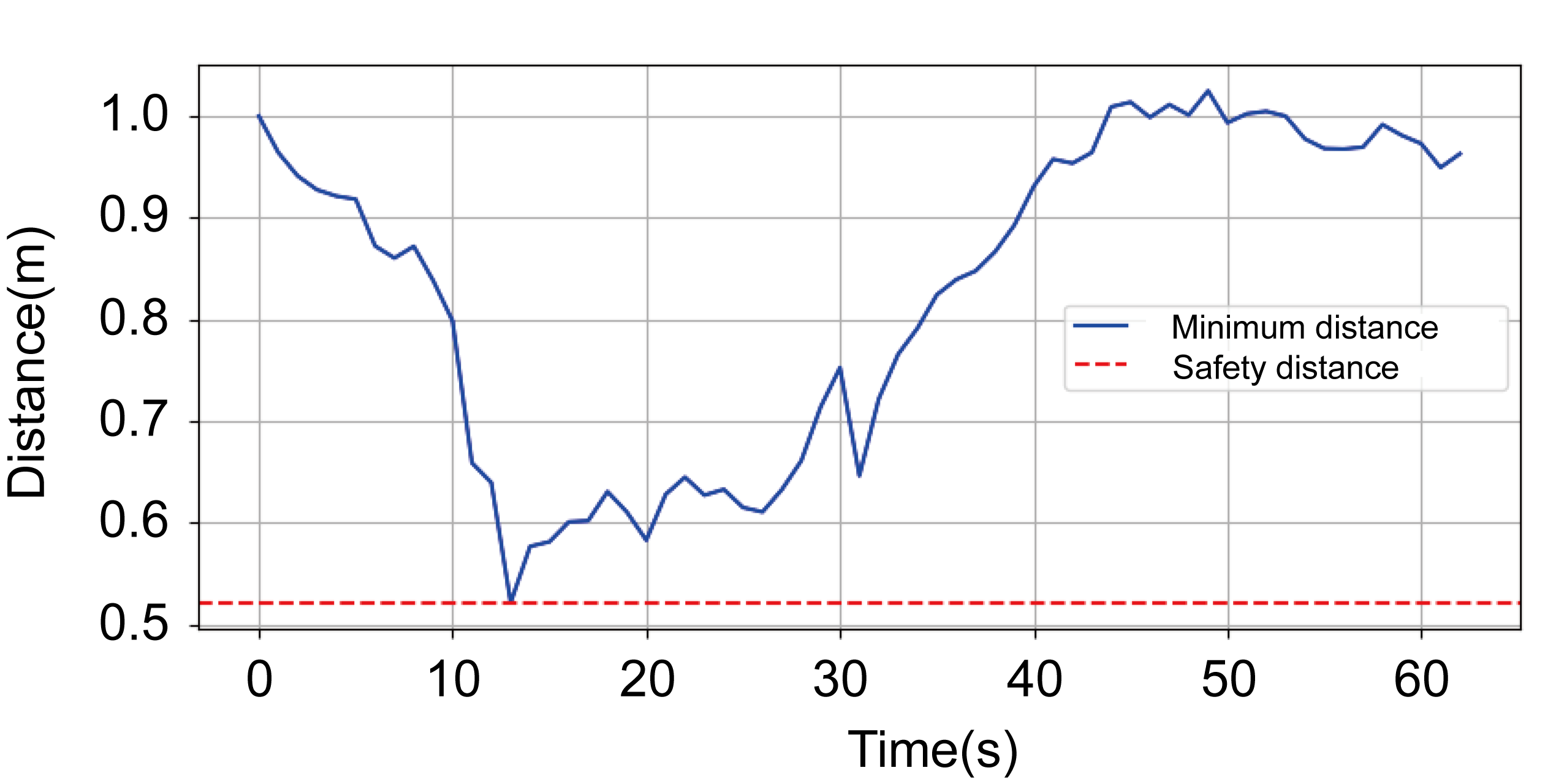}
    \caption{Minimum pairwise distance between agents over time.}
    \label{fig:minimum_distance_3AOI}
\end{figure}

\begin{figure}[htbp]
    \centering
    \includegraphics[width=0.45\textwidth]{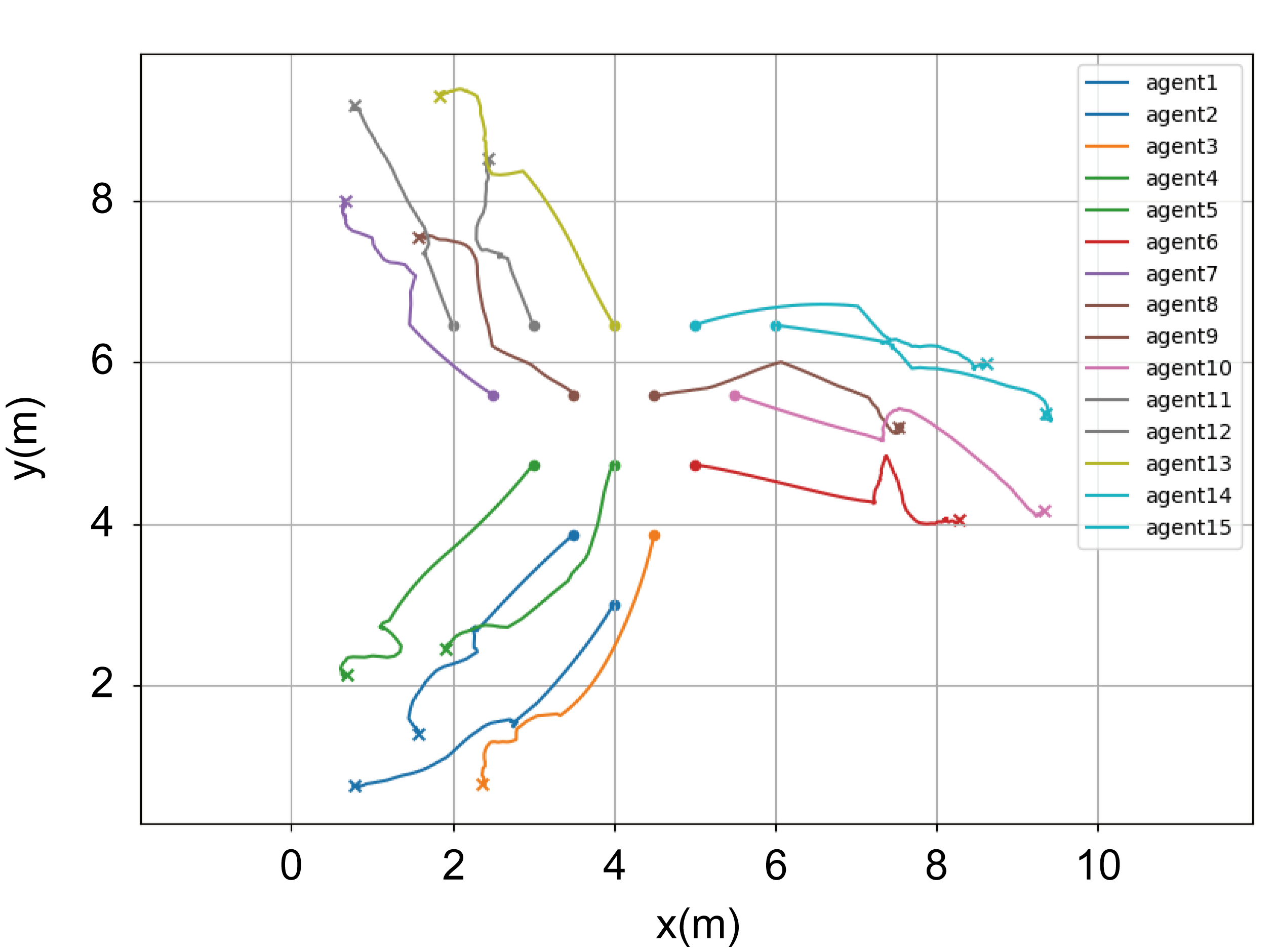}
    \caption{Complete agent trajectories during the simulation.}
    \label{fig:agent_paths_3AOI}
\end{figure}

\begin{figure}[htbp]
    \centering
    \subfloat[]{\includegraphics[width=0.50\textwidth]{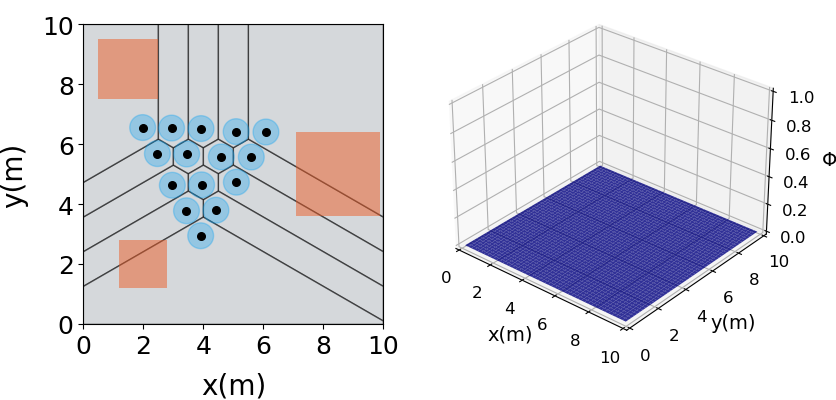}}
    \hfill
    \subfloat[]{\includegraphics[width=0.50\textwidth]{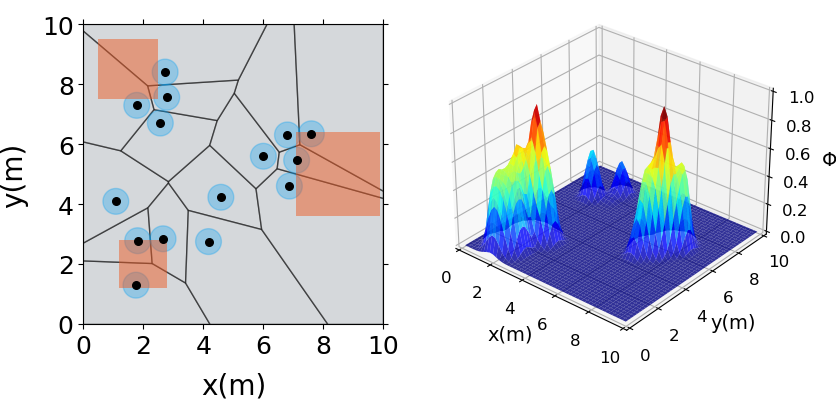}}\\
    \subfloat[]{\includegraphics[width=0.50\textwidth]{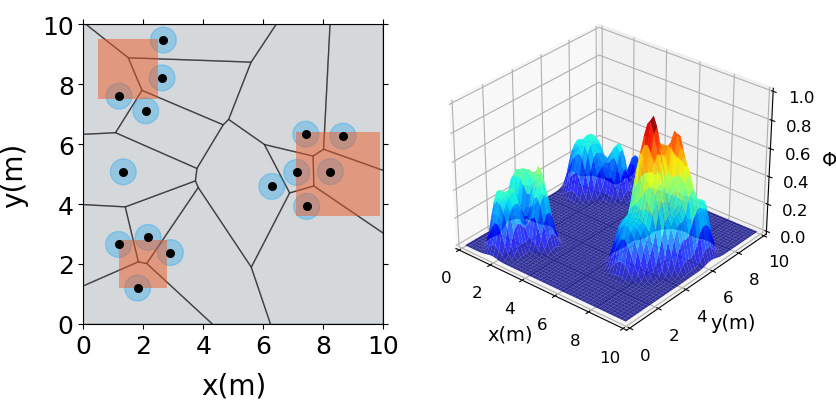}}
    \hfill
    \subfloat[]{\includegraphics[width=0.50\textwidth]{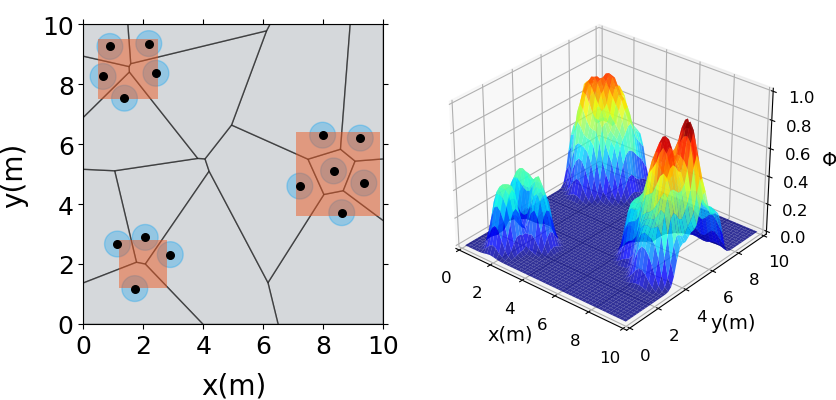}}
    \caption{Snapshots of the simulation process at different time steps: (a) $t = 0\,\mathrm{s}$, (b) $t = 14\,\mathrm{s}$, (c) $t = 23\,\mathrm{s}$, and (d) $t = 64\,\mathrm{s}$.
}
    \label{fig:coverage_snapshots_3AOI_dif}
\end{figure}

\begin{figure}[htbp]
    \centering
    \includegraphics[width=0.45\textwidth]{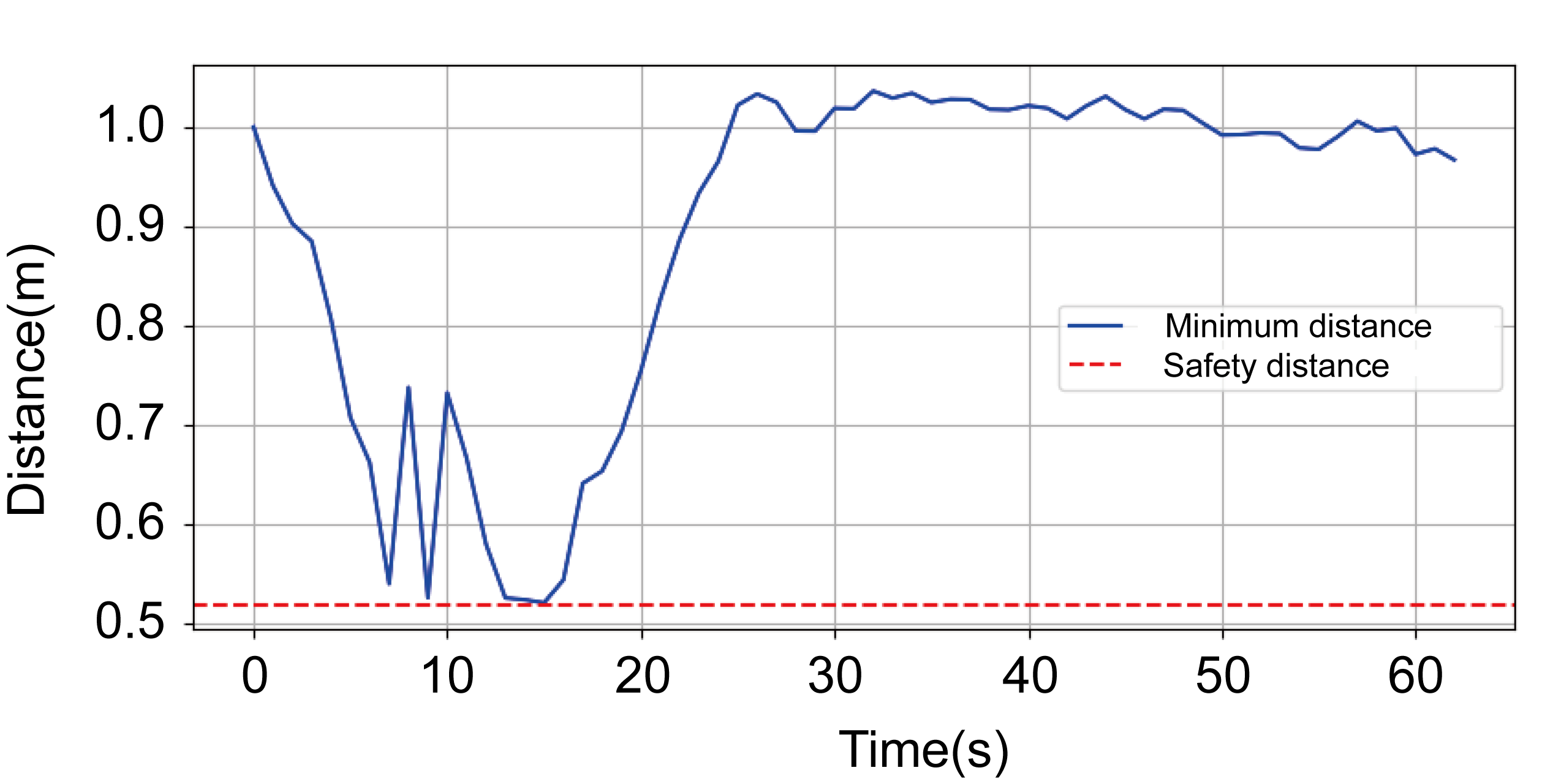}
    \caption{Minimum pairwise distance between agents over time.}
    \label{fig:minimum_distance_3AOI_dif}
\end{figure}

\begin{figure}[htbp]
    \centering
    \includegraphics[width=0.45\textwidth]{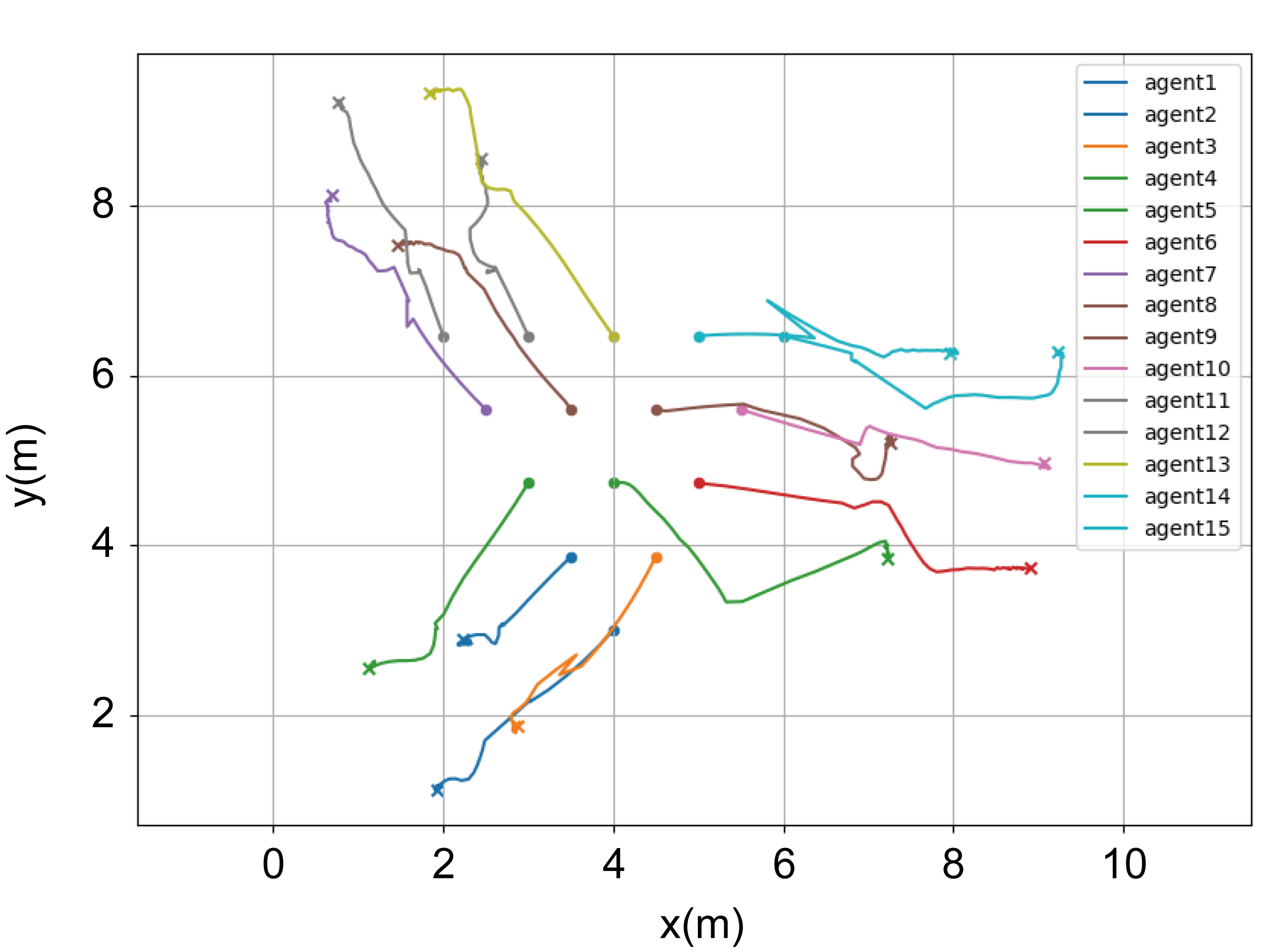}
    \caption{Complete agent trajectories during the simulation.}
    \label{fig:agent_paths_3AOI_dif}
\end{figure}

    To validate the feasibility of the proposed algorithm, we developed a simulation using Python. In the simulation environment, red square regions are designated as target areas to be covered by the multi-agent system. Initially, $N=10$ agents are uniformly arranged in a 2-row by 5-column grid, distributed across a gray workspace area without any prior knowledge of the target areas’ locations. Since our method adopts a centralized control architecture, all control commands are generated by a centralized controller and transmitted to each agent. Therefore, the total number of agents in the system is known in advance. The black dots in the figures represent the agents, and the red circles surrounding them indicate their sensor ranges. The black lines represent the boundaries of the Voronoi cells. The left panel of each figure shows the agents' positions and their sensing ranges, as illustrated in Fig.~\ref{fig:coverage_snapshots_2AOI}(a).
    
    Once an agent detects a target area within its sensing range, it generates a density function centered at the detected location to attract nearby agents, as illustrated in Fig.~\ref{fig:coverage_snapshots_2AOI}(b). As more agents are drawn in and detect one of the target areas, they generate additional density functions at their respective detection points. These density functions are superimposed, reinforcing the attraction toward each target, as shown in Fig.~\ref{fig:coverage_snapshots_2AOI}(c). The right panel in each figure visualizes the density function generated based on the agents’ sensor detections.

    The simulation concludes when all agents are able to sense at least one of the two target areas. At this point, CVT ensures that the agents are evenly distributed across the two square target areas, as shown in Fig.~\ref{fig:coverage_snapshots_2AOI}(d), while CBF maintains safe distances between agents with a minimum safety radius of $d_{\min} = 0.64$~m.
    
    The results confirm that the proposed method can successfully guide all agents toward spatially separated unknown regions and achieve uniform coverage. The density-driven CVT mechanism plays a key role in attracting agents even when only partial information is initially available. Meanwhile, Fig.~\ref{fig:minimum_distance_2AOI} verifies that the CBF effectively prevents collisions, as the minimum pairwise distance consistently remains above the safety threshold. The full trajectories shown in Fig.~\ref{fig:agent_paths_2AOI} demonstrate coordinated and smooth convergence without oscillation or redundant clustering, highlighting the robustness and scalability of the method for covering multiple disjoint areas.

To further validate the feasibility of the proposed algorithm, we extended the simulation by increasing the number of target areas to three and the number of agents to $N=15$. In this setup, three red square regions are defined as target areas that need to be covered by the multi-agent system. At the initial stage, all agents are arranged in the shape of a regular equilateral triangle within the workspace, without any prior knowledge of the target areas’ positions. The black dots represent the agents, and the red circles surrounding them indicate their sensor ranges, as shown in Fig.\ref{fig:coverage_snapshots_3AOI}(a).

When an agent detects one of the target areas within its sensor range, it generates a density function centered at the detection point to attract nearby agents, as shown in Fig.\ref{fig:coverage_snapshots_3AOI}(b). As more agents approach and detect the target areas, they generate additional density functions, which are superimposed and enhance the overall attraction to each target area, as illustrated in Fig.\ref{fig:coverage_snapshots_3AOI}(c).

The simulation concludes when all agents are able to sense at least one of the three target areas within their sensor ranges. At this point, CVT ensures that the agents are evenly distributed across the three target areas, as shown in Fig.~\ref{fig:coverage_snapshots_3AOI}(d). The CBF continues to maintain safe distances between agents with a minimum safety radius of $d_{\min} = 0.52$~m, preventing overlaps in their sensor ranges and ensuring collision-free motion.

Fig.\ref{fig:minimum_distance_3AOI} shows the minimum pairwise distance between agents over time (blue line), alongside the safety threshold defined by the CBF (red dashed line). The persistent separation between the two curves confirms the effectiveness of the safety mechanism. Furthermore, Fig.\ref{fig:agent_paths_3AOI} illustrates the complete trajectories of all agents, validating the proposed method’s ability to guide the swarm toward all three target areas in a distributed, efficient, and safe manner.

This scenario demonstrates that the proposed centrally controlled method scales effectively with both the number of agents and the number of unknown target areas. The agents self-organize into three coverage groups through globally coordinated behavior, even without predefined region segmentation. Fig.~\ref{fig:minimum_distance_3AOI} confirms the safety guarantee enforced by the CBF, while Fig.~\ref{fig:agent_paths_3AOI} illustrates smooth and efficient motion patterns. The absence of redundant clustering or idle agents further validates the robustness and effectiveness of our centralized strategy in complex multi-target coverage tasks.

To further examine the adaptability of the algorithm to heterogeneous coverage demands, we conducted an additional simulation involving $N=15$ agents and three target areas of different sizes. As shown in Fig.~\ref{fig:coverage_snapshots_3AOI_dif}(a), all agents begin by dispersing from an equilateral triangle formation without any prior knowledge of the target areas’ locations.

As agents detect target areas of different sizes, the resulting density functions naturally reflect their relative importance. Larger targets generate more overlapping density peaks, thereby attracting more agents (Fig.~\ref{fig:coverage_snapshots_3AOI_dif}(b,c)). Eventually, agents redistribute in proportion to the area size, as shown in Fig.~\ref{fig:coverage_snapshots_3AOI_dif}(d), with the largest region attracting 6 agents, the smallest attracting 4, and the remaining attracting 5.

These results demonstrate the algorithm’s capacity for size-aware distribution without requiring explicit knowledge of target geometry. The density function implicitly encodes task demand, while CVT ensures efficient partitioning, and CBF maintains safety constraints with a minimum safety radius of $d_{\min} = 0.52$~m. The safety performance is further validated by the minimum distance profile in Fig.~\ref{fig:minimum_distance_3AOI_dif}, which remains consistently above the threshold. The agent trajectories in Fig.~\ref{fig:agent_paths_3AOI_dif} highlight decentralized and balanced convergence behavior across multiple scales.

\section{Conclusion} \label{conclusion}
This paper proposed a novel goal assignment-based coverage control algorithm designed for multi-agent systems operating without prior knowledge of the target area. The agents are capable of autonomously detecting, converging upon, and efficiently covering the region of interest. By integrating CVT with a dynamically constructed density function, agents are guided to optimally distribute themselves across the detected region. In addition, the incorporation of CBF ensures safety by preventing inter-agent collisions and maintaining non-overlapping sensor coverage, thereby improving exploration efficiency.

Simulation results demonstrate the feasibility of the proposed method, showing that agents can actively search for and autonomously cover the target area. However, since the approach falls under goal assignment strategies, it inherently relies on centralized coordination, which may limit its scalability in large-scale multi-agent systems. Furthermore, as a centralized algorithm, the failure or dropout of an agent can negatively affect the system’s performance, leading to reduced coverage in the portion of the target area assigned to that agent. In future work, we aim to develop a decentralized version of the algorithm to enhance the robustness of the system and explore its application to real-world scenarios such as cooperative transportation.


\end{document}